\documentstyle[12pt]{article}

\begin{document}

\def\boxit#1{\vbox{\hrule\hbox{\vrule\kern3pt
         \vbox{\kern3pt#1\kern3pt}\kern3pt\vrule}\hrule}}
\setbox1=\vbox{\hsize 33pc  \centerline{ \bf Institut f\"ur Theoretische
Physik der Universit\"at Regensburg} }
$$\boxit{\boxit{\box1}}$$
\vskip0.8truecm

\hrule width 15.0 cm  \vskip1pt \hrule width 15.0 cm height1pt
\vskip3pt
April 1994  \hfill{TPR-94-7}
\vskip3pt\hrule  width 15.0 cm height1pt  \vskip1pt  \hrule width 15.0 cm
\vskip 2cm

\renewcommand{\thefootnote}{\fnsymbol{footnote}}
\centerline{\LARGE Nucleon Structure Functions from}
\vskip 0.4cm
\centerline{\LARGE Relativistic Constituent Quarks
\footnote{\noindent
Work supported by BMFT grant 06 OR 735. \\
\hspace*{0.5cm} To appear in Physics Letters B.}}
\vskip 1.3cm
\centerline{W.Melnitchouk and W.Weise}
\vskip 0.5cm
\centerline{Institut f\"ur Theoretische Physik}
\centerline{Universit\"at Regensburg}
\centerline{D-93040 Regensburg, Germany}
\vskip 1.5cm
\centerline{ABSTRACT}
\begin{center}
\begin{minipage}[t]{12.7cm}
{\normalsize
We investigate deep inelastic lepton scattering from the nucleon
within a constituent quark picture, in which the
internal structure of constituent quarks is modeled
by meson and diquark dressing.
In a covariant framework this structure leads to a breakdown
of the factorization necessary for a convolution formulation.
We perform our analysis in time-ordered perturbation theory
in the infinite momentum frame where factorization
of subprocesses is automatic.
Numerical results are compared with recent data on
valence quark distributions in the proton.    
}
\end{minipage}
\end{center}
\vskip1cm
\setcounter{footnote}{0}
\renewcommand{\thefootnote}{\arabic{footnote}}

\newpage


Many attempts have been made to link the parton picture
of the nucleon, which is evident in high-energy
processes such as deep inelastic scattering (DIS),
to the familiar quark models of hadrons which successfully describe
much of the low-energy phenomenology.
Most calculations of nucleon quark distributions
thus far can be grouped into two broad
categories --- those which use various bag or soliton models
of the nucleon \cite{BAGS}, and those based on the
non-relativistic quark model \cite{CONST}.
More recently, an alternative formulation in terms of
relativistic quark--nucleon vertex functions \cite{VFNS}
has led to some encouraging results.
The advantage here is that one deals directly
with covariant Feynman diagrams; on the other hand,
without decomposing the vertex functions into
explicit nucleon--quark and nucleon--antiquark components
the connection between the vertex functions and quark
and antiquark probability distributions in a general
reference frame is not obvious.

In this paper we suggest how one may combine the advantages
of some of the existing approaches
--- maintaining a close link with the constituent
quark picture of the nucleon while at the same time using
a formalism which is fully relativistic.
We adopt the traditional view that constituent quarks (CQs)
themselves have a non-trivial internal structure, which can
in principle be resolved by a high-energy probe.
In the impulse approximation, DIS from the nucleon can then
be viewed as a two-step process, in terms of
virtual photon--CQ and CQ--nucleon subprocesses.

The framework which we use is based on
time-ordered perturbation theory in the infinite momentum
frame (IMF) \cite{WEIN}.
Since here only on-mass-shell particles are involved,
factorization of subprocesses is automatic.
Therefore one may express the momentum distribution of quarks
inside a nucleon ($q_N$) as a convolution of a constituent quark
momentum distribution in a nucleon ($f_{Q/N}$) and the structure
function of a constituent quark ($q_Q$).
Factorization is generally {\em not} possible in a covariant
framework where particles are off-mass-shell.
In any frame other than the IMF one must include in addition
diagrams with particles moving backwards in time (``$Z$-graphs''),
which will introduce corrections to any convolution formula.

To describe DIS from a CQ we consider a simple dynamical model,
namely the Nambu \& Jona-Lasinio (NJL) model \cite{NJL,WW},
in which gluonic degrees of freedom are integrated out
and absorbed into an effective point-like quark interaction,
characterized by an effective coupling constant, $g$.
Since the NJL model is constructed to approximate QCD in the
low-energy and long-wavelength limit, the CQ
distributions will be those at a scale
relevant to the model itself
($Q^2 = Q_0^2 \stackrel{<}{\sim}$ 1 GeV$^2$).
Explicit perturbative gluonic degrees of freedom reappear
when the calculated distributions are evolved
to the higher $Q^2$ appropriate for the DIS region.
This procedure may be questionable since one
assumes the validity of perturbative QCD in a region
where higher order corrections are not negligible.
On the other hand, the next-to-leading order analysis of
Ref.\cite{GRV} (which uses $Q_0^2 \sim$ (0.4 GeV)$^2$)
indicates 10-20\% effects at intermediate $x$ when two-loop
radiative corrections are included.

In the NJL model the CQ structure function is determined
by the forward quark--quark ($Q$--$Q$) scattering amplitude
\cite{BJM}, which at zeroth-order in $g$
gives rise to the process depicted in Fig.1(a).
In this case the CQ structure function is a
delta-function, and the resulting nucleon quark
distribution is given by the function $f^{(S)}_{Q/N}$:
\begin{eqnarray}
q_N^{{\rm bare}(S)}(x)
= \int_x^1 {dy \over y}\ \int_0^{\infty}\ dp_T^2\
  f^{(S)}_{Q/N}(y,p_T)\  \delta(1 - x_Q)
= \int_0^{\infty}\ dp_T^2\ f^{(S)}_{Q/N}(x,p_T)         \label{qNbare}
\end{eqnarray}
where $x = Q^2 / 2 P\cdot q$ is the Bjorken scaling variable,
$y = p \cdot q / P \cdot q$ is the fraction of the nucleon's momentum
carried by the constituent quark,
and $x_Q = x/y$ is the fraction
of the constituent quark's momentum carried by the struck quark.
In Eq.(\ref{qNbare}) the superscript $S$ denotes a quark distribution
associated with a spectator diquark system in a spin $S=0$ or $S=1$ state.
The label ``bare'' refers to a constituent quark that is not
dressed by $Q$--$Q$ interactions.

Beyond the trivial order in $g$,
in the pole approximation the $t$-channel amplitude
is well approximated (at least in the pseudoscalar sector)
by using meson propagators,
and by diquark propagators in the $s$-channel.
This naturally gives rise to a dressing of the constituent quark
by meson and diquark ``clouds'', Figs.1(b) and (c),
which has the effect of softening the hard, $\delta$-function,
``bare'' quark distribution in Eq.(\ref{qNbare}).
The $s$-channel diquark-exchange process
in Fig.1(c) in fact generates an antiquark component of
the constituent quark, which in turn provides a simple
mechanism by which a non-perturbative sea can arise
in the nucleon.
The dressing corrections will be most relevant in the region
of small $x$, and higher order corrections involving additional
ladders of exchanged mesons and diquarks will not be relevant
for valence quark distributions, on which we focus in the present
work.

Having calculated their structure within a model, we assume
that CQs embedded in a nucleon have some
characteristic momentum distribution,
which in the present approach is determined by the relativistic
quark--diquark--nucleon ($QDN$) vertex function,
$\Phi^{(S)}_{QDN}$.
The Dirac structure and momentum dependence of the $QDN$ vertex
can in principle be very complicated, and our treatment at this
point is phenomenological.
Following earlier work \cite{VFNS,MST} we approximate the
vertex function by two structures,
one for scalar and one for pseudovector diquarks.
A more elaborate treatment of vertex functions, for example by
solving covariant Faddeev or Bethe-Salpeter equations \cite{BS},
is beyond the scope of the present work, but will be dealt with
elsewhere \cite{MW}.

The CQ structure function and the $QDN$ vertex
function then determine the valence quark distribution
in the nucleon:
\begin{eqnarray}
q_N^{{\rm val}(S)}(x)
&=& Z\ q_N^{{\rm bare}(S)}(x)\
 +\ \delta^{(M)} q_N^{(S)}(x)\
 -\ \delta^{(D)} \bar q_N^{(S)}(x)              \label{qNval}
\end{eqnarray}
where the three terms correspond to the ``bare'', meson-exchange,
and diquark-exchange processes in Figs.1(a), (b) and (c),
respectively.
At order $g^2$ one could also include the process where
a diquark is struck, which would require modeling in addition
the structure function of an
off-shell diquark \cite{SST}.
In the present analysis we do not consider these processes explicitly,
but rather assume that they can be absorbed into the effective
$QDN$ vertex in Fig.1(a) (see below).
The consequences for the valence quark distributions will be
negligible provided the normalization constant $Z$ is chosen to
ensure overall baryon number conservation.
Direct scattering from a pion, however, would enter only
at ${\cal O} (g^4)$.

For a scalar ($S=0$) diquark we consider a vertex
$\propto (\not\!p\ - \not\!p')/M$, which gives:
\begin{eqnarray}
f_{Q/N}^{(0)}(x,p_T)
&=& { 1 \over 16 \pi^2 }
    { \left| \Phi^{(0)}_{QDN}(x,p_T) \right|^2
      \over  x (1-x)\ (M^2 - s_{QD})^2 }
    { 1 \over M^2 }
    \left( 2 M m_Q\ (m_D^2 + m_Q^2)\
    \right.                                             \nonumber\\
& & \hspace*{-1.5cm}
    \left.
         +\ 2 P\cdot p\ (m_Q^2 - m_D^2)\
         -\ 4 M m_Q p \cdot p'\
         -\ 4 m_Q^2 P \cdot p'\
         +\ 4 P\cdot p'\ p \cdot p'
    \right)                                             \label{fQN0}
\end{eqnarray}
where $m_{Q(D)} \simeq 0.4 (0.6)$ GeV is the constituent quark
(scalar diquark) mass,
and $ s_{QD} \equiv (p + p')^2\
        =\ (m_Q^2 + p_T^2) / x\
        +\ (m_D^2 + p_T^2) / (1-x) $
is the squared mass of the $Q-D$ system.
To generate sufficiently hard quark distributions with a
structure $\propto I$ \cite{VFNS,MST,SST}
one would need more singular behavior in $p_T$ of the
vertex function $\Phi^{(0)}_{QDN}$
than the simple dipole form used below.

Any realistic model of the nucleon must also incorporate
pseudovector diquarks.
We model the quark distribution for an $S=1$ diquark with
a vertex $\gamma_5 \gamma_{\alpha}$, which gives:
\begin{eqnarray}
f_{Q/N}^{(1)}(x,p_T)
= { 1 \over 16 \pi^2 }
    { \left| \Phi^{(1)}_{QDN}(x,p_T) \right|^2
      \over x (1-x)\ (M^2 - s_{QD})^2 }
    \left( 6 M m_Q
         + 2 P \cdot p
         + {4 P \cdot p'\ p\cdot p' \over m_D^2}
    \right).                                           \label{fQN1}
\end{eqnarray}
The different $S=0$ and $S=1$ vertices, as well as a
larger pseudovector diquark mass ($m_D \simeq 0.8$ GeV),
explicitly break spin-flavor SU(4) symmetry.
This not only provides a simple explanation of the $\Delta-N$
mass difference, but also the large-$x$ $F_{2n}/F_{2p}$ ratio
\cite{CT}.

Contributions to nucleon quark distributions from DIS off dressed
CQs quarks can also be formulated as convolutions, similar
to Eq.(\ref{qNbare}).
This is only possible, however, because of the on-mass-shell
condition for interacting particles in the time-ordered theory.
In a covariant formalism the non-trivial Dirac structure of the
$\gamma^*$--$Q$ scattering amplitude and the off-shell dependence of
the $QDN$ vertex render factorization invalid.
This can be demonstrated by writing the CQ
structure function as:
\begin{eqnarray}
q_Q(x_Q)
&=& {\rm Tr}
    \left[ (\not\!p + m_Q)\ \widehat q_Q(x_Q,p^2)
    \right]                                     \label{qQdef}
\end{eqnarray}
where the operator
$\widehat q_Q\ = I\ \widehat{q}_0\
               + \not\!p\ \widehat{q}_1\
               + \not\!q\ \widehat{q}_2$
describes DIS from an off-shell CQ.
In terms of the coefficients $\widehat{q}_{0\cdots 2}$
the CQ structure function can then be written:
\begin{eqnarray}
q_Q(x_Q)
&=& 4 m_Q\      \widehat{q}_0\
 +\ 4 m_Q^2\    \widehat{q}_1\
 +\ 4 p\cdot q\ \widehat{q}_2.                  \label{qQ}
\end{eqnarray}
For a point-like quark, in the scaling limit one has
$\widehat{q}_2 \ne 0$, while $\widehat{q}_{0,1} = 0$.
On the other hand, for a CQ with internal structure
all three coefficients can be non-zero.
Indeed, dressing by either a meson or diquark cloud
gives $\widehat{q}_0 = {\cal O}(1)$,\
      $\widehat{q}_1 = {\cal O}(1)$,\
      $\widehat{q}_2 = {\cal O}(1/q^2)$.
In terms of $\widehat q_Q$ the quark distribution
in the nucleon is:
\begin{eqnarray}
q_N(x)
&=& \int d^4p\ \
{\rm Tr} \left[
    \left( I {\cal A}_0(p^2)
         + \gamma_{\alpha} {\cal A}_1^{\alpha}(p^2)
    \right) \widehat{q}_Q(x_Q,p^2)
         \right]\                               \label{qNcov}
\end{eqnarray}
where the functions ${\cal A}_{0,1}$
represent the off-shell $Q$--$N$ interaction.
Because of the scaling properties of $\widehat q_{0\cdots 2}$,
the trace in Eq.(\ref{qNcov}) will not be proportional to
the CQ structure function $q_Q$ in Eq.(\ref{qQ}).
Therefore Eq.(\ref{qNcov}) can only be written in convolution
form if the functions ${\cal A}_{0,1}$ are proportional,
however this will not be the case either unless the on-mass-shell
limit is taken for these.
This result is analogous to that for an off-shell nucleon
bound inside a nucleus \cite{MST}.
However, unlike in the nuclear case where nucleon off-shell
corrections are not overwhelming \cite{MST,KPW},
neglect of quark off-shell effects in the nucleon can
lead to significant errors.
On the other hand, because of the on-mass-shell kinematics
in the time-ordered approach,
the $QDN$ interaction factorizes, allowing
$\delta^{(M)} q_N^{(S)}$ and $\delta^{(D)} \bar q_N^{(S)}$
in Eq.(\ref{qNval}) to be expressed as two-dimensional
convolutions as in Eq.(\ref{qNbare}).

For the meson dressing process in Fig.1(b) we restrict
ourselves to pions.
Dressing of constituent quarks by higher mass mesons
(e.g. $\rho, \omega$, etc.) can be considered,
but contributions from these will be suppressed
due to the larger meson masses.
The contribution to $q_N$ from a constituent quark
dressed by a pion is given by:
\begin{eqnarray}
\delta^{(\pi)} q^{(S)}_N(x)
&=& \int_x^1 {dy \over y}\
    \int_0^{\infty} dp_T^2\ f_{Q/N}^{(S)}(y,p_T)\
    \int d^2k_T\ q_Q(x_Q,k_T,p_T)                       \label{qpi}
\end{eqnarray}
where $f_{Q/N}^{(S)}$ is given by Eqs.(\ref{fQN0}) and (\ref{fQN1}),
and
\begin{eqnarray}
q_Q(x_Q,k_T,p_T)
&=& { g_{Q\pi}^2 \over 16 \pi^3 }
    {|\Phi_{QQ\pi}(x_Q,k_T,p_T)|^2
     \over x_Q (1-x_Q) (m_Q^2 - s_{Q\pi})^2}            \nonumber\\
& & \hspace*{0cm} \times
    \left( {k_T^2 + m_Q^2 (1-x_Q)^2 \over x_Q}\
         +\ x_Q\ p_T^2\ -\ 2\ {\bf p}_T \cdot {\bf k}_T
    \right)
\end{eqnarray}
represents the $\gamma^*$--(dressed) CQ interaction.
Here $g_{Q\pi} = m_Q/f_{\pi} \approx 4$ is the $Q$--$\pi$ coupling
constant \cite{WW}, as given by the Goldberger-Treiman relation,
and $s_{Q\pi}
\equiv (k~+~k')^2\
 =\ (m_Q^2~+~k_T^2) / x_Q\
 +\ (m_{\pi}^2 + ({\bf p}_T~-~{\bf k}_T)^2) / (1-x_Q)\
 -\ p_T^2$.
In terms of $q_Q(x_Q,k_T,p_T)$ the structure function of
a (free) constituent quark at rest (Eq.(\ref{qQdef}))
is: \ \ $q_Q(x_Q) = \int_0^{\infty} dk_T^2\ q_Q(x_Q,k_T,0)$.

The presence of transverse momentum $p_T$ in {\em both}
$f_{Q/N}$ and $q_Q$ in Eq.(\ref{qpi}) means that
a simple one-dimensional convolution in $y$ alone
cannot be valid unless the $p_T$-dependence
in $q_Q$ is neglected.
This turns out to be a rather bad approximation, however,
as the $p_T=0$ results are found to differ by up to 30-40\%
from the full, $p_T$-dependent calculation of the
pion-dressed contributions \cite{MW}.

In Eq.(\ref{qpi}) we assume that the integral of $q_Q$ over
$k_T$ is regularized by a sharp cut-off $\Lambda$, and consequently
parametrize the $QQ\pi$ vertex function by a theta-function,\ \
$ \Phi_{QQ\pi}
= \theta \left( k_T^2 - (1-x_Q) \Lambda^2 \right)$,
with $\Lambda \approx 0.9$ GeV \cite{WW}
characteristic of a typical chiral symmetry breaking scale.
The contribution of the diquark-exchange process in
Fig.1(c) to the nucleon antiquark distribution is
similar to the $\pi$ contribution in Eq.(\ref{qpi}),\ \
$\delta^{(\pi)} q^{(S)}_N(x) \rightarrow
 \delta^{(D)} \bar q_N^{(S)}(x)$,
but with $m_{\pi} \rightarrow m_D$
and $g_{Q\pi} \rightarrow g_{QD} \approx 4$ taken from
the NJL model.

The only additional input necessary for the numerical evaluation
of $q_N^{{\rm val}(S)}(x)$ is the $QDN$ vertex
function $\Phi^{(S)}_{QDN}$.
For this we take a simple dipole ansatz,
$ \Phi_{QDN}^{(S)}(x,p_T)\
\propto\ \left( \left( \Lambda_S^2 - m_Q^2 \right)
              / \left( \Lambda_S^2 - t     \right)
         \right)^2$,\
where
$t \equiv - \left( p_T^2 + x (m_D^2 - (1-x) M^2) \right)/$ $(1~-~x)$.
The cut-offs $\Lambda_S$ are chosen to reflect the presumably
smaller radius of a scalar diquark compared with an $S$=1
diquark:\ $\Lambda_0 = 1.5$ GeV and $\Lambda_1 = 1.2$ GeV.
For comparison, we also consider an
$s_{QD}$-dependent dipole vertex function,
$\Phi^{(S)}_{QDN} \propto (s_{QD} + \Lambda_S^2)^{-2}$,
and an exponential type,
$\Phi^{(S)}_{QDN} \propto
\exp \left( (M^2 - s_{QD})/\Lambda_S^2 \right)$.
These choices are motivated by the resulting symmetry of
quark and diquark probability distributions under the
interchange $y \leftrightarrow 1-y$ and
$m_Q \leftrightarrow m_D$, which one would need to
satisfy if direct scattering from
virtual diquarks was included.
Since, as mentioned above, these contributions are absorbed
into the $QDN$ vertex function in Fig.1(a),
this symmetry need not be explicit, and $t$-dependent
vertex functions may be used.
The main difference between these forms arises at small $x$,
where the $s_{QD}$-dependent functions give somewhat smaller
distributions.
The reason for this is the $1/x$ factor in $s_{QD}$,
which at small $x$ serves to suppress the quark distributions,
which themselves depend on inverse powers of $s_{QD}$.
A full investigation of the effects of various
vertex functions will be made elsewhere \cite{MW}.

In Figs.2 and 3 we plot the resulting
valence quark distribution in the proton,
$x (u_V + d_V) = 3 x (q_N^{{\rm val}(0)}
                    + q_N^{{\rm val}(1)}) / 2$,
as well as the $d_V / u_V$ ratio.
The calculated distribution (dashed curve) has been evolved from
$Q_0^2 \approx$ (0.32 GeV)$^2$ to $Q^2 = 5$ GeV$^2$ (solid curve)
using a contour integration procedure similar to that outlined
in Ref.\cite{GRV} to invert the Mellin transform.
The shaded region represents the range of several
parametrizations of world data \cite{PARAM}.
Clearly the agreement with the data is very good.
Even better fits to the data could be obtained by modifying
some of the parameters in the analysis, such as diquark
masses, form factor cut-offs, or the input scale $Q_0^2$.
Our choices, however, are motivated by physical arguments
and previous calculations \cite{WW},
and we feel the results are encouraging enough to warrant
further investigation.

In conclusion, we see that a relativistic constituent quark
model formulated in the IMF can lead to reasonable
valence quark distributions in the nucleon.
Independent of any model approximations,
the time-ordered approach used here has
the advantage that factorization of subprocesses
is automatic.
In a covariant formulation the off-shell structure of
constituent quarks renders the convolution approach
inappropriate.
The present work can also be extended to model the sea
of the nucleon.
Already the diquark-exchange process in Fig.1(c) contributes
some 20-30\% of the nucleon antiquark distribution
at $Q^2 \sim 5$ GeV$^2$ \cite{MW}.
Higher order processes involving multi-pion and multi-diquark
exchanges, with correspondingly heavier intermediate states,
will further enhance the $x \rightarrow 0$ region.
Finally, the spin structure \cite{SPIN} of relativistic constituent
quarks can be studied within the same framework; the details of
this will appear in a forthcoming publication \cite{MW}.

\vspace*{0.3cm}
W.M. would like to thank H.Meyer, A.W.Schreiber, K.Steininger and A.W.Thomas
for discussions related to some of the points addressed in this paper.
This work was supported by the BMFT grant 06 OR 735.


\newpage

{\bf Figure captions.}

1. Deep inelastic scattering from (a) a constituent quark
   in the nucleon;
   (b) a constituent quark dressed by pions;
   (c) a constituent quark dressed by diquarks.
   The photon four-momentum is $q$, and
   $P$, $p$ and $p'$ are momenta of the nucleon,
   constituent quark and spectator diquark, respectively.

2. Calculated valence $x(u_V + d_V)$ quark distribution in the
   proton (dashed),
   and evolved to $Q^2 = 5$ GeV$^2$ (solid).
   The shaded region envelopes the parametrizations of world data
   from Ref.\protect\cite{PARAM}.

3. Ratio of proton $d_V$ to $u_V$ quark distributions.
   The curves are as in Fig.2.

\end{document}